\documentclass[times,twocolumn]{aastex631}
\usepackage{graphicx}

\shorttitle{Gravitational Instability around an FUor object}
\shortauthors{Weber et al.}

\begin{document}

\title{Spirals and clumps in V960 Mon: signs of planet formation via gravitational instability around an FU Ori star?}

\author[0000-0002-3354-6654]{Philipp Weber}
\email{philipppweber@gmail.com}
\author[0000-0003-2953-755X]{Sebasti{\'a}n P{\'e}rez}
\affiliation{Departamento de Física, Universidad de Santiago de Chile, Av. Victor Jara 3659, Santiago, Chile.}
\affiliation{Millennium Nucleus on Young Exoplanets and their Moons (YEMS), Chile.}
\affiliation{Center for Interdisciplinary Research in Astrophysics and Space Exploration (CIRAS), Universidad de Santiago de Chile, Chile.}

\author[0000-0002-5903-8316]{Alice Zurlo}
\affiliation{Millennium Nucleus on Young Exoplanets and their Moons (YEMS), Chile.}
\affiliation{Núcleo de Astronomía, Facultad de Ingeniería y Ciencias, Universidad Diego Portales, Av. Ejercito 441, Santiago, Chile.}
\affiliation{Escuela de Ingenier\'ia Industrial, Facultad de Ingenier\'ia y Ciencias, Universidad Diego Portales, Av. Ejercito 441, Santiago, Chile.}

\author[0000-0002-1575-680X]{James Miley}
\affiliation{Joint ALMA Observatory, Alonso de Córdova 3107, Vitacura, Santiago 763-0355, Chile.}
\affiliation{National Astronomical Observatory of Japan (NAOJ), Los Abedules 3085, Office 701, Vitacura, Santiago, Chile.}

\author[0000-0001-5073-2849]{Antonio Hales}
\affiliation{National Radio Astronomy Observatory, 520 Edgemont Road, Charlottesville, VA 22903-2475, United States of America.}

\author[0000-0002-2828-1153	]{Lucas Cieza}
\affiliation{Millennium Nucleus on Young Exoplanets and their Moons (YEMS), Chile.}
\affiliation{Núcleo de Astronomía, Facultad de Ingeniería y Ciencias, Universidad Diego Portales, Av. Ejercito 441, Santiago, Chile.}

\author[0000-0002-7939-377X]{David Principe}
\affiliation{MIT Kavli Institute for Astrophysics and Space Research, 77 Massachusetts Avenue, Cambridge, MA 02139, USA.}

\author[0000-0003-0564-8167]{Miguel C{\'a}rcamo}
\affiliation{Millennium Nucleus on Young Exoplanets and their Moons (YEMS), Chile.}
\affiliation{Center for Interdisciplinary Research in Astrophysics and Space Exploration (CIRAS), Universidad de Santiago de Chile, Chile.}
\affiliation{University of Santiago of Chile (USACH), Faculty of Engineering, Computer Engineering Department, Chile.}

\author[0000-0002-4266-0643]{Antonio Garufi}
\affiliation{INAF, Osservatorio Astrofisico di Arcetri, Largo Enrico Fermi 5, 50125 Firenze, Italy.}

\author[0000-0001-7157-6275]{{\'Agnes} K{\'o}sp{\'a}l}
\affiliation{Konkoly Observatory, Research Centre for Astronomy and Earth Sciences, E\"otv\"os Lor\'and Research Network (ELKH), Konkoly-Thege Mikl\'os \'ut 15-17, 1121 Budapest, Hungary.}
\affiliation{CSFK, MTA Centre of Excellence, Konkoly Thege Mikl\'os \'ut 15-17, 1121 Budapest, Hungary.}
\affiliation{ELTE E\"otv\"os Lor\'and University, Institute of Physics, P\'azm\'any P\'eter s\'et\'any 1/A, 1117 Budapest, Hungary.}
\affiliation{Max Planck Institute for Astronomy, K\"onigstuhl 17, 69117 Heidelberg, Germany.}

\author[0000-0001-9248-7546]{Michihiro Takami}
\affiliation{Institute of Astronomy and Astrophysics, Academia Sinica, 11F of Astronomy-Mathematics Building, No.1, Sec. 4, Roosevelt Rd, Taipei 10617, Taiwan, R.O.C.}
\author[0000-0002-3138-8250]{Joel Kastner}
\affiliation{School of Physics \& Astronomy, Rochester Institute of Technology, 1 Lomb Memorial Dr., Rochester, NY 14623,
USA.}
\author[0000-0003-3616-6822]{Zhaohuan Zhu}
\affiliation{Department of Physics and Astronomy, University of Nevada, Las Vegas, 4505 S. Maryland Pkwy, Las Vegas, NV, 89154, USA.}
\affiliation{Nevada Center for Astrophysics, University of Nevada, Las Vegas, Las Vegas, NV 89154, USA.}
\author[0000-0001-5058-695X]{Jonathan Williams}
\affiliation{Institute for Astronomy, University of Hawai’i at Manoa, Honolulu, HI 96822, USA.}

\begin{abstract}
The formation of giant planets has traditionally been divided into two pathways: core accretion and gravitational instability.
However, in recent years, gravitational instability has become less favored, primarily due to the scarcity of observations of fragmented protoplanetary disks around young stars and low occurrence rate of massive planets on very wide orbits.
In this study, we present a SPHERE/IRDIS polarized light observation of the young outbursting object V960~Mon.
The image reveals a vast structure of intricately shaped scattered light with several spiral arms.
This finding motivated a re-analysis of archival ALMA 1.3$\,$mm data acquired just two years after the onset of the outburst of V960~Mon.
In these data, we discover several clumps of continuum emission aligned along a spiral arm that coincides with the scattered light structure.
We interpret the localized emission as fragments formed from a spiral arm under gravitational collapse.
Estimating the mass of solids within these clumps to be of several Earth masses, we suggest this observation to be the first evidence of gravitational instability occurring on planetary scales.
This study discusses the significance of this finding for planet formation and its potential connection with the outbursting state of V960~Mon.
\end{abstract}

\section{Introduction}\label{sec:intro}
The core accretion scenario, characterized by the continuous growth of dust particles followed by runaway gas accretion, is the prevailing formation scenario for gas giant planets \citep{Pollack1996}. However, when it comes to directly imaged giant planets and brown dwarfs found at significant distances from their host stars, most models face difficulties in producing sufficiently massive objects within the expected lifetimes of gaseous disks \citep[e.g.][]{Emsenhuber2021}. As a result, gravitational instability (GI) emerges as a prominent alternative for planet formation in these regions \citep[see][for a review]{Kratter2016}, as it is believed to operate in the outer protoplanetary disk. 

The underlying physical mechanisms that drive the formation of massive gas giant planets through gravitational instability in circumstellar disks may also be responsible for the observed episodic accretion events in young stellar objects \citep{Armitage2001, Fischer2022}. According to this model, protostars undergo intermittent and transient, yet highly efficient accretion phases, known as ``FUor events'' (named after the prototype event observed in FU~Orionis). Those events are characterized by a significant increase in {accretion} luminosity occurring over an annual timescale.
FUor objects pose an excellent laboratory to study planet formation, as the sudden increase in brightness of the {central source} illuminates and heats the surrounding environment.
This has been exploited in the past for the first direct detection and characterization of the water ice line \citep{Cieza2016,Tobin2023}.
For an extensive discussion of FUor objects, existing observational evidence, and its theoretical interpretations we refer to dedicated review articles \citep{Audard2014,Fischer2022}.

V960~Mon {(2MASS~J06593158-0405277)} is a bona fide FUor object \citep[][]{Connelley2018,Cruz2023} that has been in an outbursting state since 2014 \citep{Maehara2014}.
The distance to the object has remained a subject of controversy over the years. Initially, \citet{Kim2004} proposed a kinematic distance of 2.3$\,$kpc based on CO observations and association with the molecular cloud S~287. However, when compared to stellar evolution models, a distance of approximately 450$\,$pc was estimated \citep{Kospal2015}. More recently, \citet{Kospal2021} used a distance of 1574$\,$pc from the Gaia~DR2 dataset. These contrasting results underscore the sensitivity of the inferred distance to the chosen method of measurement.

\citet{Kospal2015} analyzed several pre-burst archival datasets across the spectrum and proposed eight companions in the local field to be pre-main sequence objects in a T Tauri stage. From this, the authors suggest that V960~Mon is not isolated. 
Most recently, \citet{Cruz2023} observed the blue- and red-shifted parts of an outflow in $^{12}$CO emission approximately along the line-of-sight. From $^{13}$CO measurements, they estimate a massive envelope of $\sim0.6\,{\rm M}_\odot$. 

In this Letter, we present and analyze prominent large-scale spirals around the FUor V960~Mon observed with SPHERE/IRDIS in polarized light.
By using archival ALMA 1.3$\,$mm continuum data we reveal dust clumps within these structures. 
We describe the observational setup in Section~\ref{sec:data}, present the results in Section~\ref{sec:results}, and discuss their implications in Section~\ref{sec:discussion}.
These observations potentially connect GI clumps to a recent FUor outburst and offer an unprecedented opportunity to characterize gravitational instability at planet-forming scales.

\section{Observation and data reduction}\label{sec:data}
We observed V960~Mon in $H$-band ($\lambda = 1.625\,$µm) in the night of 2016 December 17 (programme-ID: 098.C-0422(B), PI: L. Cieza) in dual-beam polarimetric
imaging mode (DPI, \citealp{deBoer2020,vanHolstein2020}) with the InfraRed Dual-band Imager and Spectrograph (IRDIS, \citealp{Dohlen2008}) of VLT/SPHERE \citep{Beuzit2019}.
The target was part of a SPHERE survey to detect the scattered light from the environments around six episodic accreting FUors and EXors, which will be presented in Zurlo et al. (in prep).
The observing conditions during operation were excellent (seeing\,$\sim$\,0\farcs4, $v_{\rm wind}$\,$\sim$\,$3.5\,{\rm m}\,{\rm s}^{-1}$, $\tau_0$\,$\sim$10\,ms).
We used the {\tt N\_ALC\_YJH\_S} coronagraph (185$\,$mas diameter, {\citealp{Carbillet2011,Guerri2011}}) centered on V960~Mon and took a total of four half-wave plate (HWP) cycles of four object frames each. Each object frame was exposed to a detector integration time (DIT) of 64$\,$s, adding up to a total integration time of 17$\,$min. Immediately before and after taking the polarimetric science frames, we took star center frames (DIT of 64$\,$s) to accurately infer the stellar position behind the coronagraph and star flux frames (DIT of 2$\,$s) to calibrate the observed intensities.

We used the IRDAP reduction pipeline \citep[IRDIS Data reduction for Accurate Polarimetry, version 1.3.4,][]{vanHolstein2020} to process the SPHERE/IRDIS data and extract the Stokes $Q$ and $U$ components of linear polarization.
The pipeline incorporates a comprehensive model of the SPHERE optical system, enabling direct correction for instrumental polarization and polarization crosstalk without relying on the data.
IRDAP automatically corrects for the reduction of the unresolved stellar polarization due to the central star.
We further subtract the unresolved polarization carried in the stellar halo of the close-in companion according to \citet[][Appendix B]{Weber2023}. The total linearly polarized intensity is then calculated as:
\begin{equation}
    PI = \sqrt{Q^2+U^2}\,.
\end{equation}
We refrain from using the conventional $Q_\varphi/U_\varphi$ representation commonly employed in polarized light imaging, as it is not suitable for capturing the polarized intensity if off-centered light sources are present within the local environment.

\section{Results}\label{sec:results}
\begin{figure}
    \centering
    \includegraphics[width=\columnwidth]{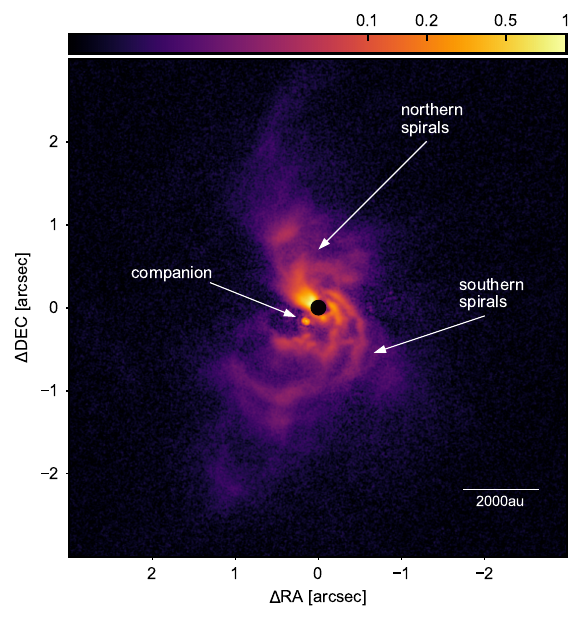}
    \caption{SPHERE/IRDIS linearly polarised intensity ($PI$) observation of V960~Mon in $H$-band. $PI$ is normalized to its maximum value (6.6$\,$mJy$\,$arcsec$^{-2}$) and on a logarithmic scale. The frame is centered on the primary, which was covered by the coronagraph displayed by a black circle. The projected scale is shown in the lower right assuming the distance of 2189$\,$pc.}
    \label{fig:VLT}
\end{figure}
Figure~\ref{fig:VLT} shows the IRDIS polarized light $H$-band image for V960~Mon. 
The frame is centered on the primary star which is concealed behind a coronagraph marked by a black circle.
The image displays a vast S-shaped structure of scattered light extending along the north-south axis.
Both the northern and southern parts are comprised of at least two adjacent spiral arms each.
Assuming the Gaia DR3 distance \citep[$d=2189 \pm 281\,$pc,][]{Gaia2022}, the projected extent of these spiral arms is several thousand au. Although we should approach the Gaia DR3 distance with caution due to the relatively large renormalized weighting unit error (RUWE) of 3.75, it is noteworthy that the Gaia DR3 distance calculated for the assumed companion UCAC4~430-024261 (also referred to as V960~Mon~N) is similar at 2606$\pm$346$\,$pc \citep{Gaia2022}, with a sufficiently small RUWE of 0.93, indicating a reliable astrometric solution.
However, it is important to interpret the projected spatial scales presented in this study while considering the controversy associated with the distance measurement of the object (as introduced in Section~\ref{sec:intro}).

We confirm a close stellar companion southeast of the coronagraph
%\citep[originally detected in][]{CarratioGaratti2015}
which appears both in polarized and unpolarized intensity.
{This object was previously detected in $J$- and $K$-band, with Pa$\beta$ and Br$\gamma$ emission at a distance of 227$\,$mas and a PA of 131.4$\,$deg \citep{CarratioGaratti2015}.
Here,} we measure the companion to be at a distance of 237$\pm4\,$mas and at a PA of 136.7$\pm1.0\,$deg with respect to the primary. Notably, this companion is co-located with the scattering structures, meaning there is no evidence of its orbit being cleared (see Fig.~\ref{fig:VLT}). {The detection of the companion in both polarized and unpolarized intensity implies the presence of small dust grains in its immediate vicinity or indicates significant scattering along the line of sight.}

\begin{figure*}
    \centering
    \includegraphics[width=\textwidth]{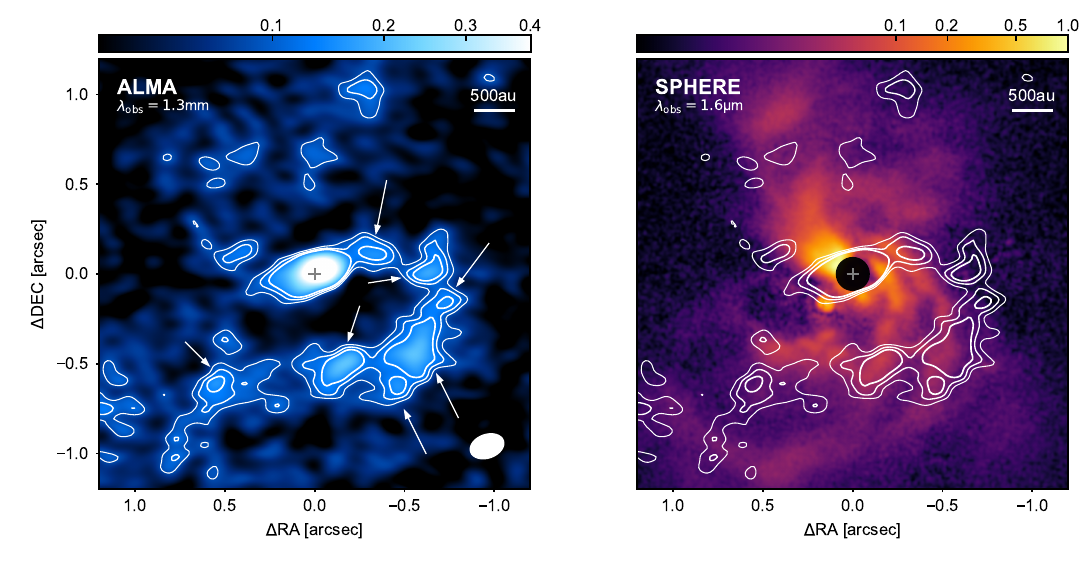}
    \caption{Comparison of ALMA band~6 continuum image in mJy$\,$beam$^{-1}$ using natural weights (left panel) with SPHERE/IRDIS polarized light image (right panel, see Fig.~\ref{fig:VLT}). Both images are overlaid with contours of the ALMA continuum at levels of 3, 4, and 5 $\sigma_{\rm rms}$ (with $\sigma_{\rm rms}=28$µJy$\,$beam$^{-1}$). The clean beam of the ALMA observation is shown in white in the bottom right corner of the left panel ($0\farcs14\times0\farcs20$).{The arrows in the left panel mark the dust clumps.}}
    \label{fig:master}
\end{figure*}
Motivated by the spectacular scattered light image seen in Fig.~\ref{fig:VLT}, we conducted a re-examination of archival ALMA band~6 data (programme-ID: 2016.1.00209.S, PI: Takami) previously published in \citet{Kospal2021}.
We describe our data reduction and the differences to \citet{Kospal2021} in Appendix~\ref{appendix:ALMAdata}, with an exploration of the imaging parameters to test for the robustness of the observed structures.

In the left panel of Figure~\ref{fig:master} we show the ALMA 1.3$\,$mm continuum image and compare it to the polarized light image in the right panel. Both panels superimpose ALMA contours at 3, 4, and 5 $\sigma_{\rm rms}$ levels (with $\sigma_{\rm rms}=28\,$µJy$\,$beam$^{-1}$). 
Remarkably, these contours reveal multiple spatially-separated fragments of continuum emission, reaching up to 7$\sigma$. They align along a clockwise-opening spiral arm originating from the primary source.   
The left panel of Fig.~\ref{fig:master} shows that these clumps roughly coincide with the southern spiral structure observed in scattered light.
We note that {although the respective observations were taken within a small temporal separation,} a perfect alignment is not expected as the scattered light image probes the illuminated surface of the structures, while the ALMA continuum image traces emission from dust in cold, dense clumps that are likely optically thick in the near-infrared (NIR). {In Appendix~\ref{appendix:spiral}, we show that the mm clumps can align with the scattered light spiral by assuming an adequate orientation for the spiral and a function for the scattering surface.}

Both the emission centered on V960~Mon and the individual clumps remain unresolved by the beam of the observation ($0\farcs14\times0\farcs20$ for natural weighting). {We note that the close companion seen in the SPHERE data does not exhibit any significant counterpart in millimeter emission.}

Assuming the 1.3$\,$mm continuum emission is optically thin, we can estimate the clumps' masses based on a given temperature and dust opacity.
Considering a typical opacity of $\kappa_{\rm 1.3mm}=2.3\,{\rm cm}^2\,{\rm g}^{-1}$ \citep{Beckwith1990} and a temperature of 50$\,$K, the clumps correspond to solid material masses ranging 
from $3$ to $10\,{\rm M}_\oplus$, which corresponds to a gas mass of 1 to 3 M$_{\rm Jup}$, assuming a gas-to-dust mass ratio of 100. 
However, it is important to note that both the opacity and temperature can significantly deviate from these values, depending on dust properties and local thermodynamics. Also, the gas-to-dust value can be significantly smaller where dust accumulates.
The calculated masses exhibit an approximate inverse dependence on the assumed temperature and show a quadratic proportionality with the assumed distance to V960~Mon. This indicates that if the previously assumed  distances were used, the inferred clump masses would be even smaller. Consequently, we argue that the continuum emission effectively traces clumping occurring at scales relevant to planet formation.

\section{Discussion}\label{sec:discussion}
\subsection{Origin of spiral structures}
Spiral structures have been observed in several protostellar systems \citep[see table~2 in][]{Bae2022}.
These spirals exhibit variations in size, number, contrast, and pitch angle, suggesting different physical origins.
Commonly discussed possibilities {in the context of protoplanetary disks} are a massive external companion \citep[e.g.][]{Dong2015b} {or vortex \citep{vanderMarel2016,Huang2019}}, gravitational instability \citep[e.g.][]{Lodato2005}, an inner binary \citep{Price2018}, a stellar fly-by \citep[e.g.][]{Clarke1993} or combinations of these processes \citep{Thies2010,Pohl2015,Meru2015}.

The complex environment surrounding V960~Mon presents a challenge for pinpointing the precise cause of the scattered light structure.
The significant measured envelope mass \citep{Cruz2023} implies a further potential association between the large-scale spirals and infalling material \citep{Lesur2015, Hennebelle2017,Kuffmeier2018}.
Another plausible explanation could be the capture of a close-by cloudlet \citep{Dullemond2019}.
Additionally, the presence of multiple objects in the immediate vicinity \citep{Kospal2021} introduces the possibility of considering them as potential candidates of a past fly-by.

\citet{Kospal2021} estimate that the mass corresponding to the unresolved emission around the primary could be as high as $0.33\,{\rm M}_\odot$ (updated from the original value of $0.17\,{\rm M}_\odot$ considering the Gaia DR3 distance). The authors calculate that a disk of such a mass around an approximately solar-mass star would be susceptible to gravitational instability beyond a critical radius. However, simulations indicate that disks undergoing gravitational instability typically cannot maintain spiral arms beyond a radius of 100$\,$au for long periods of time, as the disk tends to fragment at larger radii \citep{Rafikov2005, Cossins2010,Zhu2012}.

\subsection{Gravitational Fragmentation}
For V960~Mon, it remains uncertain which of the preceding formation scenarios accurately describes the environment surrounding the dust clumps. So far, Keplerian rotation was only detected in the optical and NIR \citep{Park2020}, where the molecular line profiles trace material much closer to the star.
We speculate that the spirals are located between the inner envelope and outer disk and refer to disk equations with reservation.

To our knowledge, the only process capable of explaining the fragmentation of a spiral arm into clumps (as witnessed in Fig.~\ref{fig:master}) is the gravitational instability.
The initial onset of gravitational instability is governed by the Toomre instability criterion, with $Q<1$, where $Q$ is defined as
\begin{equation}
    Q \equiv \frac{c_{\rm s}\Omega}{\pi G \Sigma}\,,
\end{equation}
with the local sound speed $c_{\rm s}$, the Keplerian frequency $\Omega$, the gravitational constant $G$ and the surface density $\Sigma$.
During gravitational instability, a disk is expected to generate large-scale spiral arms \citep{Zhu2012} strikingly similar to those seen in the scattered light around V960~Mon in Fig.~\ref{fig:VLT}. Those spirals are expected to induce shock waves throughout the disk, which heat the disk material and regulate or potentially prevent further gravitational collapse.
The ultimate fate of the gravitational collapse hinges on the efficiency of the disk material in radiating its thermal energy, directly relating to the cooling timescale \citep{Gammie2001,Rafikov2005}.

The theoretical/numerical works of \citet{Takahashi2016} and \citet{Brucy2021} have predicted that if cooling is efficient enough, fragmentation of a spiral arm can occur as the second stage of a two-step gravitational instability process.
This is characterized by an adjusted instability criterion of $Q\lesssim0.6$ within the spiral arm.

\subsection{Fate of clumps}
The detection of clumps within the ALMA band~6 observation of V960~Mon marks a significant milestone by providing the first concrete evidence of a fragmenting spiral arm.
This discovery indicates the involvement of gravitational instability in the formation of {planetary-mass} clumps and the evolutionary processes occurring within protoplanetary disks, at least in certain instances. 

What will happen to these clumps in the long run? Numerous studies have examined the fate of fragments resulting from gravitational instability. A comprehensive overview can be found in \citet{Kratter2016}. In this context, several key concepts should be highlighted.

One possibility is that the clumps disintegrate shortly after their formation. This can occur due to tidal interactions with other clumps, rapid encounters with spiral structures, or insufficient cooling associated with tidal destruction.

Furthermore, \citet{Vorobyov2015} discovered that clump interactions can lead to rapid accretion onto the central star, potentially causing an accretion outburst. As a result, a secondary clump may be ejected into the interstellar medium. These ejected clumps may be precursors of free-floating planets \citep{Sumi2011} or brown dwarfs \citep{Basu2012}.

Moreover, \citet{Zhu2012} conducted numerical simulations demonstrating that the fates of the clumps depend on the migration speed, cooling, and accretion efficiency. Slow migration leads to the formation of a massive companion opening a gap, while fast migration leads to tidal destruction. \citet{Nayakshin2010} showed that in the tidal destruction scenario, the outer, more volatile components of the clumps are sequentially stripped away, and can ultimately leave a solid core in the inner region (a process labeled tidal downsizing).

Finally, the interest in the formation of planetary cores through gravitational instability has been rekindled with the inclusion of solid material in the process of gravitational instability \citep{Baehr2022}. The authors found that under conditions specific to different dust sizes, overdensities can collapse and survive to give rise to planetary embryos. 
This holds the potential to significantly reduce the timeline of planet formation and offers a promising explanation for the detection of planetary signposts within the outer disks of even very young systems \citep{Baehr2023}.

\subsection{FUor event}
Additional evidence supporting the presence of GI can be attributed to V960~Mon's classification as a bona fide FUor object. It has long been anticipated that FUor objects would serve as promising candidates for detecting indications of GI in their surrounding disks.
Several studies have demonstrated that episodic accretion events can be triggered by the interplay between GI and MRI   \citep{Armitage2001,Zhu2009,Martin2011}. \citet{Vorobyov2005} further established that the inward spiraling clumps resulting from GI can lead to episodes of intense accretion, mirroring the observed behavior of FUor objects.

{The environments of most FUor objects look disrupted when observed the NIR scattered light (\citealp{Liu2016}, \citealp{Takami2018}, Zurlo et al. in prep.). The mm-continuum, however, had thus far exhibited no signatures of any dynamical perturbances (see V883~Ori, \citealp{Cieza2016}; V900~Mon, \citealp{Takami2019}; FU Orionis, \citealp{Perez2020}; HBC~494, \citealp{Nogueira2023}).} 
This raises an immediate question regarding the persistence of {the fragmented structures for mm grains}. According to \citet{Klahr2020}, the timescale for a dust clump's contraction is estimated to be $\tau_{\rm c}\sim (9{S_{t}\Omega})^{-1}$, where ${S}_{t}$ represents the Stokes number representative of the dust's dynamical behavior.
{For dust particles of $S_t=1$ at 100$\,$au around solar-mass stars, this contraction timescale would be as short as $\sim20\,$years. After contraction, the clumps might not be detectable any longer.}

\section{Conclusions \& Implications}
We presented a polarized light image obtained with the VLT/SPHERE instrument, revealing the intricate system surrounding the FUor object V960~Mon. The image exhibits remarkable spiral arms extending over hundreds of astronomical units. Building upon these findings, we reanalyzed archival ALMA data, which provided deeper insights into the structure of the spirals, uncovering clumpy features located slightly offset with respect to the scattered light emission. We quantified the properties of these clumps and estimated their dust and gas masses, marking the first detection of clumps in the planetary mass regime. The observed characteristics of V960~Mon closely resemble the predictions from simulations of gravitational instability, underscoring the suitability of FUor objects as laboratories for studying planet formation.

The timing of the ALMA observations is particularly significant, as they were conducted merely two years after the stellar outburst. However, it is crucial to acknowledge that the FU~Ori phenomenon might involve multiple triggering mechanisms, with gravitational instability-induced fragmentation representing just one possibility among several others. To gain further insights, a follow-up of FUor outbursts detected by the Legacy Survey of Space and Time (LSST) using the Vera C. Rubin telescope, specifically targeting millimeter clumps, could help establish the prevalence of such features around FUor sources in the early stages of outburst events.

\begin{acknowledgments}
We thank the anonymous referee for a constructive report. We thank Cornelis Dullemond for useful discussions on infall from the environment and gravitational instability acting on large scales. 
P.W. acknowledges support from FONDECYT grant 3220399.
This work was funded by ANID -- Millennium Science Initiative Program -- Center Code NCN2021\_080.
S.P. acknowledges support from FONDECYT Regular grant 1231663.
%Lucas:
L.C. acknowledges support from FONDECYT Regular grant 1211656.
%Agnes:
This project has received funding from the European Research Council (ERC) under the European Union's Horizon 2020 research and innovation programme under grant agreement No 716155 (SACCRED).
%Puelche
This work made use of the Puelche cluster hosted at CIRAS/USACH.
%VLT
The work is based on observations collected at the European Southern Observatory under ESO programme 098.C-0422.
%ALMA
This paper makes use of the following ALMA data: ADS/JAO.ALMA\#2016.1.00209.S. ALMA is a partnership of ESO (representing its member states), NSF (USA) and NINS (Japan), together with NRC (Canada), NSTC and ASIAA (Taiwan), and KASI (Republic of Korea), in cooperation with the Republic of Chile. The Joint ALMA Observatory is operated by ESO, AUI/NRAO and NAOJ. The National Radio Astronomy Observatory is a facility of the National Science Foundation operated under cooperative agreement by Associated Universities, Inc.
%Gaia
This work has made use of data from the European Space Agency (ESA) mission {\it Gaia} (\url{https://www.cosmos.esa.int/gaia}), processed by the {\it Gaia} Data Processing and Analysis Consortium (DPAC, \url{https://www.cosmos.esa.int/web/gaia/dpac/consortium}). Funding for the DPAC has been provided by national institutions, in particular the institutions participating in the {\it Gaia} Multilateral Agreement.
\end{acknowledgments}

\vspace{5mm}
\facilities{VLT, ALMA, Gaia}
\dataav{The images are available as fits files at \href{https://github.com/yemsnucleus/V960_Mon_ApJL}{https://github.com/yemsnucleus/V960\_Mon\_ApJL}}
\software{
This work has made use of the IRDAP-pipeline \citep[version 1.3.4,][]{vanHolstein2020} for the processing of SPHERE/IRDIS data and the CASA software \citep[version~5.6.1-8 and version~6,][]{CASA} for the processing of the ALMA data.
For self-calibration, we used an automated module for the ALMA Pipeline (Tobin et al., in prep.). For alternative imaging, we used the GPUVMEM package \citep{Carcamo2018}.
We used IPython \citep{ipython}, NumPy \citep{numpy} and Matplotlib \citep{Matplotlib} for data analysis and creating figures. 
}

\bibliographystyle{aasjournal}

\begin{appendix}
\section{ALMA data reduction}\label{appendix:ALMAdata}
\begin{figure*}
    \centering    
    \includegraphics[width=\textwidth]{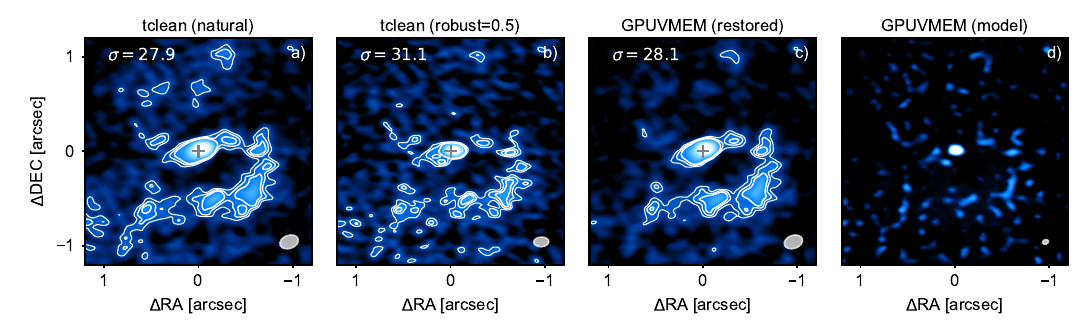}
    \caption{{Applying synthesis imaging reconstruction techniques to the ALMA band~6 data. a) and b) show the {\tt tclean} reduction, a) using natural weighting (same as in Fig.~\ref{fig:master}) and b) using briggs-weighting with robust=0.5. The resulting clean beam sizes are $0\farcs14\times0\farcs20$ and $0\farcs10\times0\farcs16$, respectively. 
    Panels c) and d) show the {\tt GPUVMEM} reconstruction technique applied to the data. Panel c) was restored using the clean beam of panel a), panel d) shows the image model.
    In all panels, the clean beam is displayed in translucent white in the bottom right corner. We measure the rms-noise, $\sigma$ for each image individually, written in the top left corner of each respective image and in units of µJy$\,{\rm beam}^{-1}$. The contours are shown at levels of 3$\sigma$, 4$\sigma$, and 5$\sigma$. The position of the central star is marked by a $+$. The {\tt GPUVMEM} model does not allow a measurement of the rms-noise.}
    }
    \label{fig:appendix}
\end{figure*}
Observations of V960 Mon were taken as part of the project 2016.1.00209.S (PI: M. Takami) and previously published in \citet{Kospal2021}.
The data were obtained using a combination of ALMA configurations; these included an extended 12-m array configuration (2017 July 27, maximum baseline 3.7$\,$km, pwv~0.47$\,$mm) and a compact configuration (2017 April 20, 460$\,$m, 2.2$\,$mm).
Two spectral windows of bandwidth 1.875$\,$GHz were dedicated to the continuum at central frequencies 216.877$\,$GHz, and 232.178$\,$GHz.
Three other spectral windows were set up for molecular line observations, each with a total bandwidth of 0.059$\,$GHz, positioned with central frequencies 230.514$\,$GHz, 220.375$\,$GHz, and 219.537$\,$GHz to cover the rotational 2--1~transitions of $^{12}$CO, $^{13}$CO, and C$^{18}$O respectively, using channel widths of 15.259$\,$kHz, 30.158$\,$kHz, and 30.158$\,$kHz.
We reduced the CO molecular line data to examine potential dynamical features evident in the moment 1 maps.
However, the analysis yielded inconclusive results due to the limited on-target observation time of only $\sim12\,$min in total.
Data calibration was carried out using the CASA pipeline in software version 5.6.1-8. Imaging and further analysis use CASA version~6. In our Letter, we present continuum imaging that combines all frequency channels devoid of line emission, employing a total aggregate bandwidth of 3.445$\,$GHz. 

{We utilized the standalone version of the automated self-calibration module for the ALMA Pipeline (Tobin et al., in prep.)\footnote{\href{https://github.com/jjtobin/auto_selfcal}{https://github.com/jjtobin/auto\_selfcal}} to perform data self-calibration. The self-calibration process was performed separately for each array configuration, with two iterations of phase-only self-calibration. Following self-calibration, the compact configuration data showed an increase in signal-to-noise ratio of 17\%. However, the more extended baseline dataset exhibited only a marginal improvement ranging from 1\% to 2\%. Subsequently, both datasets were combined by concatenation to generate the final composite image.}
 
The imaging presented within this Letter in Fig.~\ref{fig:master} applies natural weighting to the visibilities, resulting in a synthesized beam size of $0\farcs14\times0\farcs20$. 
To provide a comparison, we also applied an alternate weighting scheme of briggs (robust=0.5) in Figure~\ref{fig:appendix}, which improved the beam size to $0\farcs10\times0\farcs16$.
However, this enhancement comes at a slight cost to the signal-to-noise ratio of the clumps. 
In Fig.~\ref{fig:appendix}, it is evident that the size of the clumps decreases as the beam size becomes smaller, indicative of their approximate point source nature.

{To ensure that the identified clumps are not artifacts resulting from the {\tt tclean} image synthesis process, we reconstructed an independent image with the maximum entropy method (MEM) using the {\tt GPUVMEM}\footnote{\href{https://github.com/miguelcarcamov/gpuvmem}{https://github.com/miguelcarcamov/gpuvmem}} package \citep{Carcamo2018}. To apply this method, we utilized a penalization factor of $\lambda=0.01$ in the entropy regularization term \citep[see][for a description of the relevant parameters]{Carcamo2018}. The MEM reconstruction is an independent and unsupervised algorithm known for its ability to provide the highest resolution while maintaining sensitivity. Since the MEM algorithms forward-model visibility measurements using nonparametric models of the sky brightness distribution following well-informed image priors, the model image provided by {\tt GPUVMEM} is a good representation of the actual sky brightness. In order to account for residuals and have an image comparable to the one produced by {\tt tclean}, we also present a "restored" version of the model, which convolves the model with the natural weighted beam and adds the residuals. In Fig.~\ref{fig:appendix}, panel c) exhibits the restored image that matches the resolution of the natural weights. Remarkably, the restored version closely resembles the {\tt tclean} image, underscoring the independence of the identified features from the image synthesis process. Fig.~\ref{fig:appendix} panel d) displays the image model that effectively enhances resolution to approximately one-third of the original scale. The model image suggests that the clumps are significantly smaller than the natural beam size, corroborated by the contours in panel b). Both panel b) and d) illustrate that the emission centered on the primary source in panel a) encompasses two additional clumps that merge with the central feature for the larger beam sizes.}

A final word about the previous publication of the ALMA data: the clumps are not prominent in the same ALMA data presented in \citet{Kospal2021} likely due to the authors' primary focus on resolving the inner disk and their use of uniform gridding in the imaging process. 
This approach compromises the signal-to-noise ratio, resulting in the faint clumps being closer to the noise level.
However, it should be noted that the clumps also exhibit partial $5\sigma$-level signals when applying uniform weights \citep[see fig.~1 in][]{Kospal2021}.
In our study, we were guided by the IRDIS observation, motivating us to prioritize the detection of subtle features. Therefore, we utilized natural weighting in our imaging process to optimize for point-source sensitivity. This decision helped us highlight and investigate the previously unnoticed fragmenting spiral in greater detail.

\section{Fitting the spiral}\label{appendix:spiral}
\begin{figure}
    \centering
    \includegraphics[width=0.8\textwidth]{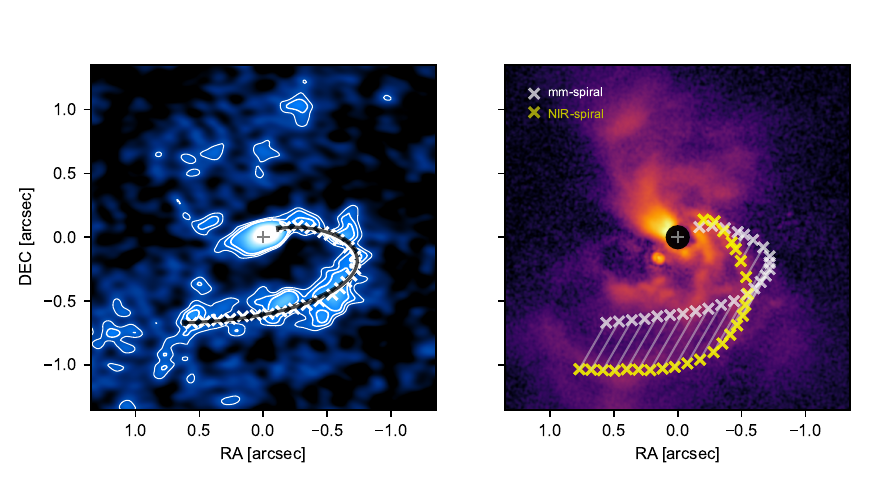}
    \caption{Fitting consistent spirals to the images: The left panel shows the ALMA data with the 3, 4, and 5$\,\sigma_{\rm rms}$ contours. The white crosses represent the data points used for tracing the spiral arm. The black curve corresponds to a function fitted to the deprojected data points, which are then re-projected onto the image plane. 
    The SPHERE observation in the right panel includes the same white crosses delineating the mm-spiral. Additionally, the yellow crosses represent the projection of the white crosses onto the scattering surface, assuming the function for the scattering height, $h_{\rm scat}$, described in the text.}
    \label{fig:spiral_fit}
\end{figure}
In Section~\ref{sec:results}, we discussed that the traced spiral arm of the mm-clumps in the ALMA image does not precisely match the spiral arm observed in the SPHERE image. We commented that this discrepancy arises due to the distinct nature of the observations. ALMA primarily captures thermal radiation emitted directly from the dust grains, while SPHERE images the surface where the dust structure becomes optically thick to incident stellar irradiation and scatters those photons to the observer. Consequently, the observed radiation originates from two different planes, and when observed under an inclination, the projection of the spiral can lead to different locations in the image plane, resulting in a incongruent visual effect.

Our objective is to investigate whether and under what conditions the mm-clumps can be aligned with the closest spiral arm observed in scattered light. To achieve this, we make the assumptions that all mm-clumps lie within a single plane and that the scattered light arises directly above the mm-spiral. The comparison is carried out in two steps.
First, we deproject the imaged ALMA spiral by assuming fixed values for inclination and PA. 
We fit an exponential spiral function of the form $\phi(r) = a\exp(br)$ to the deprojected ALMA data points, where $r$ represents the deprojected radial distance to the center and $\phi$ denotes the azimuthal coordinate in the deprojected plane, increasing in the counterclockwise direction. 
The $R^2$-value of the fit is best for an inclination of $61^\circ$ and a PA of 285$^\circ$.
However, it is important to note that there were several other combinations of inclination and PA that resulted in nearly equally satisfactory outcomes.
The spiral parameters at this deprojection yield an amplitude of $a=356^\circ$ and a rate of change of $b=-4.5\times10^{-4}\,$au$^{-1}$.
In the second step, we explore the parameter space of the scattering surface function, $h_{\rm scat} = h_0\times(r/r_0)^\psi$, to identify the values that align the mm-spiral with the scattered light spiral. We set $r_0=500\,$au and find that the best results are obtained for $h_0=0.15$ and $\psi=0.5$. Figure~\ref{fig:spiral_fit} shows the resulting fit to the spirals.
\end{appendix}

\end{document}